%% file: main.tex
\documentclass[fleqn,usenatbib]{mnras}

\usepackage{newtxtext,newtxmath}

\usepackage[T1]{fontenc}
\usepackage{ae,aecompl}

\usepackage{graphicx}	
\usepackage{amsmath}	
\usepackage{amssymb}	
\usepackage[usenames,dvipsnames]{xcolor} 
\usepackage[normalem]{ulem} 
\usepackage{hyperref}
\usepackage{array}
\usepackage{booktabs} 
\usepackage{multirow}  

\newcommand{\eagle}{\mbox{\sc{Eagle}}}
\newcommand{\flares}{\mbox{\sc Flares}}
\newcommand{\synthesizer}{\mbox{\tt Synthesizer}}
\newcommand{\prospector}{\mbox{\tt Prospector}}

\defcitealias{Weibel_2025}{W25}

\title[Early Massive Quiescent Galaxies]{The Nature of High-Redshift Massive Quiescent Galaxies - Searching for RUBIES-UDS-QG-z7 in FLARES}

\author[Turner et al.]{Jack C. Turner$^{1}$\thanks{E-mail: jt458@sussex.ac.uk}, 
Will J. Roper$^{1}$, 
Aswin P. Vijayan$^{1}$, 
Sophie L. Newman$^{2}$, 
Stephen M. Wilkins$^{1,3}$ 
\newauthor
Christopher C. Lovell$^{4,5,2}$,
Shihong Liao$^{6}$, 
Louise T. C. Seeyave$^{1}$
\\
$^{1}$Astronomy Centre, University of Sussex, Falmer, Brighton BN1 9QH, UK\\
$^{2}$Institute of Cosmology and Gravitation, University of Portsmouth, Burnaby Road, Portsmouth PO1 3FX, UK\\
$^{3}$Institute of Space Sciences and Astronomy, University of Malta, Msida MSD 2080, Malta \\
$^{4}$Kavli Institute for Cosmology, Madingley Road, Cambridge CB3 0HA, UK\\
$^{5}$Institute of Astronomy, Madingley Road, Cambridge CB3 0HA, GB\\
$^{6}$Key Laboratory for Computational Astrophysics, National Astronomical Observatories, Chinese Academy of Sciences, Beijing 100101, China
}

\date{Accepted XXX. Received YYY; in original form ZZZ}

\pubyear{2026}

\begin{document}
\label{firstpage}
\pagerange{\pageref{firstpage}--\pageref{lastpage}}
\maketitle

\begin{abstract}
RUBIES-UDS-QG-z7 (RQG) is the earliest massive quiescent galaxy identified to date, inferred to have formed its abundant stellar mass in a single burst that ceases rapidly before $z\sim8$. An object of such extreme nature challenges our understanding of galaxy formation, requiring rapid growth and quenching mechanisms only $0.6 \ \rm{Gyr}$ after the Big Bang and implying number densities $2 \ \rm{dex}$ higher than currently predicted by simulations. We use synthetic observables to identify analogous systems within the First Light And Reionisation Epoch Simulations (\flares) and find two massive galaxies ($M_{\ast}>10^{9} \ \mathrm{M_{\odot}}$) dominated by rapidly quenched bursts. One of these demonstrates excellent agreement with the inferred physical properties of RQG and implies a number density of analogous systems $\log_{10}(\mathrm{N_{Q}} \ / \ \mathrm{Mpc}^{-3}) = -7.92^{+0.52}_{-0.76}$. Beyond demonstrating that the current \flares\ model is capable of producing RQG-like systems, these analogues provide a laboratory within which to study the underlying physics. Their active galactic nuclei (AGN) heat and expel gas, inducing rapid quenching and preventing timely rejuvenation. This causes above-average chemical enrichment at a given stellar mass, with super solar levels predicted for RQG. These metallicities are underestimated by spectral energy distribution fitting and we show that $\alpha$-enhancement cannot be solely responsible. Degeneracies with age and dust attenuation appear the more likely causes. Tensions between observed and simulated number densities can be alleviated in part by considering systematics, but adjustments to AGN feedback, such as allowing super-Eddington accretion rates, may be required for full agreement.
\end{abstract}




\input Sections/1_intro
\input Sections/2_data
\input Sections/3_analogues
\input Sections/4_physics
\input Sections/5_conclusion

\section*{Acknowledgements}

We thank the authors of \cite{Weibel_2025} for sharing their \prospector\ code and associated models.

We thank the \eagle\, team for their efforts in developing the \eagle\, subgrid model. 

JCT was supported by a Science and Technology Facilities
Council (STFC) studentship ST/X508822/1.

This work used the DiRAC@Durham facility managed by the Institute for Computational Cosmology on behalf of the STFC DiRAC HPC Facility (www.dirac.ac.uk). The equipment was funded by BEIS capital funding via STFC capital grants ST/K00042X/1, ST/P002293/1, ST/R002371/1 and ST/S002502/1, Durham University and STFC operations grant ST/R000832/1. DiRAC is part of the National e-Infrastructure. 

The RQG spectrum was retrieved from the Dawn JWST Archive (DJA). DJA is an initiative of the Cosmic Dawn Center, which is funded by the Danish National Research Foundation under grant DNRF140.

We also wish to acknowledge the following open source software packages used in the analysis: \textsc{Astropy} \citep{Astropy_2022}, \textsc{CMasher} \citep{van-der-Velden_2020}, \textsc{h5py}, \textsc{Matplotlib} \citep{Hunter_2007}, \textsc{Numpy} \citep{Harris_2020} and \textsc{Scipy} \citep{Virtanen_2020}.

We list here the roles and contributions of the authors according to the Contributor Roles Taxonomy (CRediT)\footnote{\url{https://credit.niso.org/}}. \textbf{Jack C. Turner}: Conceptualisation, Data Curation, Methodology, Investigation, Formal Analysis, Visualisation, Writing - original draft. \textbf{Will J. Roper, Aswin P. Vijayan, Sophie L. Newman}: Conceptualisation, Data Curation, Methodology, Writing - review \& editing. \textbf{Stephen M. Wilkins}: Conceptualisation, Supervision, Writing - review \& editing. \textbf{Christopher C. Lovell}: Methodology, Writing - review \& editing. \textbf{Shihong Liao, Louise T. C. Seeyave}: Writing - review \& editing.


\section*{Data Availability}

\flares\ data, including the photometry generated for this work, is available upon reasonable request to the corresponding author. Code specific to this work is publicly available at \url{https://github.com/jackcturner/flares_rubies_comp}. Further information regarding \flares\ can be found at \url{https://flaresimulations.github.io}.



\bibliographystyle{mnras}
\bibliography{flares, flares-rubies}

\appendix
\input Sections/appendix_1
\input Sections/appendix_2

\bsp
\label{lastpage}
\end{document}

%% file: Sections/1_intro.tex
\section{Introduction}\label{sec:intro}

Understanding the formation and evolution of galaxies in the early Universe is central to interpreting the underlying astrophysics and cosmology driving its evolution. The \emph{Hubble Space Telescope} consistently pushed the redshift frontier to earlier cosmic times \citep{Bunker_2003, Coe_2013, Oesch_2016}, and alongside Keck and \emph{Spitzer} identified unique systems where specific physics could be probed and used to inform theoretical models \citep{Treu_1998, Dunlop_2004, Makarov_2016}. One such sub population are `quiescent' or `passive' galaxies, which are characterised by low star formation rate (SFR) to stellar mass ratios, also referred to as the specific star formation rate (sSFR). These are known to be commonplace at low-redshift, as the gradual consumption of available cold gas leads to a decrease in the cosmic star formation rate density \citep[CSFRD;][]{Madau_2014}. Local examples have been studied extensively by ground-based programmes, which tie their evolution to the dense cluster environments \citep{Vulcani_2010, Darvish_2017, Hewitt_2025} that only begin forming at $z \lesssim 2$ \citep{White_1991, Overzier_2016, Bocquet_2024}. Objects at these redshifts can be well resolved and used to constrain quenching mechanisms such as ram pressure stripping \citep{Simpson_2018, Werle_2025}, virial shocks  \citep{Dekel_2006, Gabor_2014b} and major mergers \citep{van-der-Wel_2009, Pawlik_2018, Wilkinson_2022} that are most prevalent in clusters.

These dense clusters form gradually through the process of hierarchical assembly, so are considerably less numerous at early times. While early protoclusters do exist \citep{Jiang_2018, Hu_2021, Morishita_2023}, we would not expect an abundance of these galaxies at $z > 2$ if environment was the only driver of quiescence. However, large samples have already been identified, including long before the peak in CSFRD \citep{Dominguez_2011, Straatman_2014, Santini_2021}. Internal mechanisms such as stellar \citep{Larson_1974, Geha_2012, Hopkins_2014} and active galactic nuclei \citep[AGN;][]{Silk_1998, Fabian_2012, Piotrowska_2021} feedback and gas depletion \citep{Feldmann_2014, Peng_2015}, must become increasingly important at this time. These processes are key to calibrating modern galaxy formation simulations to reproduce observed relations \citep{Choi_2018, Pandya_2020, Rosito_2021}. It is therefore essential to study these systems in detail wherever they might exist, but observatories responsible for the majority of quiescent galaxy detections only probe the rest-frame optical continuum and Balmer/4000\AA\ break at $z<4$. With the former probing recent star formation \citep{Conroy_2013, Iyer_2025} and the latter commonly used as an indicator of age \citep{Bruzual_1983, Poggianti_1997} and sensitive to the star formation history \citep[SFH;][]{Wilkins_2023b}, the discovery potential for quiescent galaxies at higher redshift is extremely limited.

The NIRCam \citep{Rieke_2023} and NIRSpec \citep{Boker_2023} instruments onboard \emph{JWST} are capable of capturing the Balmer break up to $z=11$ and $z=13$ respectively, exceeding this redshift limit with ease. As a result, there are now significant samples of quiescent galaxies at $4 < z < 6$ \citep{Valentino_2023, Russell_2024, Long_2024, Baker_2025c, Merlin_2025}, which allow simulations to be constrained by statistically robust distribution functions \citep{FLARES-VIII, Baker_2025a, Chaikin_2025, Seeyave_2025}. There have also been a number of intriguing smaller passive samples, including individual galaxies identified at $z>7$ \citep{Looser_2024, Weibel_2025, Baker_2025b}, which challenge galaxy formation simulations to enact quenching mechanisms only $\sim 700 \ \rm{Myr}$ after the Big Bang. Likewise, quiescent galaxies inferred to have formed at these these redshifts have been identified at later times \citep{Carnall_2024, de-Graaff_2024, Glazebrook_2024}. While not as significant as large samples, these galaxies facilitate alternative approaches. By fitting their star formation histories, we can constrain when their stellar populations formed and compare to the predicted limits from astrophysical theories and even cosmological models \citep{Lovell_2022, Boylan-Kolchin_2023, Carnall_2024}. Simply investigating whether current theory can produce comparable systems is a valuable model constraint \citep{Flores_2022, Qin_2023}. Alternatively, simulations can be used to qualitatively infer the properties of galaxies of particular interest, serving as a more focused counterpart to simulation-based inference \citep{Iglesias-Navarro_2024, Iglesias-Navarro_2025}. Constrained Milky-Way and local group simulations are a common example of inferring the properties of specific galaxies \citep{Bignone_2019, Wempe_2024}, but this approach has also been applied to more distant objects selected on apparent uniqueness or unexpected nature \citep{Moreno_2022, Filipp_2023}. 

A clear candidate for such an analysis is \mbox{RUBIES-UDS-QG-z7}, described in detail by \citet[][hereafter W25]{Weibel_2025} using spectroscopy taken as part of the RUBIES NIRSpec programme \citep{de-Graaff_2025}. We refer to this object as the RUBIES Quiescent Galaxy (RQG) herein. RQG was identified from photometry as a high priority target for spectroscopic follow-up due to its bright apparent magnitude and red spectral energy distribution (SED). Clear Lyman and Balmer breaks in the resulting spectrum placed strong constraints on the redshift of $z_{\rm{RQG}} = 7.29 \pm 0.01$, and SED fitting reveals it to be both massive and quiescent. This makes RQG the earliest such galaxy identified to date. Such an object poses a potential challenge for galaxy formation models, with \citetalias{Weibel_2025} showing a factor $\sim150$ discrepancy with the number density predicted by \cite{FLARES-VIII} at $z=7$. Projections from smaller periodic volumes also underestimate the densities, even under the most conservative assumptions. While concerning, these comparisons are heavily dependent on systematic choices, which are not necessarily consistent between simulations and observations. In addition, large Poisson and cosmic variance uncertainties are introduced by low number statistics and small sky areas respectively, further limiting the constraining power. Its inferred physical properties may be more remarkable than this tension, having formed its extreme stellar mass $\log_{10}(\mathrm{M_{\ast}}/\mathrm{M_{\odot}}) = 10.23 \pm 0.04$ in an intense burst lasting $\lesssim 200 \ \rm{Myr}$, preceding a rapid quenching that holds until observation. The inferred metallicity is only $\sim 0.1 \ \mathrm{Z}_{\odot}$, which is hard to explain given its stellar mass and particularly abrupt star formation history. \citetalias{Weibel_2025} posit that high levels of $\alpha$-enhancement \citep{Vazdekis_2015, Byrne_2023} could bias this measurement. A second modelling approach that attempts to account for this effect suggests a super solar metallicity, but both models return comparable $\chi^{2}$. The levels of chemical enrichment and $\alpha$-enhancement in RQG therefore remain an open question. The evolutionary pathways of such a galaxy only $\sim 600 \ \rm{Myr}$ after the Big Bang are unclear and could be interpreted as an unphysical product of SED fitting uncertainties \citep{Suess_2022, Haskell_2024, Bellstedt_2025, Nanayakkara_2025}. Regardless, RQG is almost certainly high-redshift, massive and quiescent, and challenges galaxy formation models to produce analogous systems. Further study to inform these models and pinpoint the underlying physical mechanisms is imperative.

In this work, we use forward modelled observables (\S\ref{sec:data}) to search for theoretical analogues of RQG in the First Light and Reionisation Epoch Simulations (\flares), with the aim of answering three key questions. First, can \flares\ produce galaxies that look like RQG photometrically, and are they massive quiescent galaxies (\S\ref{sec:analogues})? If so, what can we infer about the evolution of RQG and its nature at the time of observation (\S\ref{sec:physics})? Using this information, can we identify the physics that may need revision to account for the difference in observed and simulated number densities? We match the \flares\ simulation suite in assuming a Planck cosmology throughout, with $\Omega_{\rm{m}} = 0.307$, $\Omega_{\Lambda} = 0.693$ and $h = 0.6777$ \citep{planck_collaboration_2014}.

%% file: Sections/2_data.tex
\section{Data}\label{sec:data}

\renewcommand{\arraystretch}{1.4}
\begin{table*}
\centering
\begin{tabular}{c|cccc}
\toprule
\multicolumn{5}{c}{\textbf{Photometry [nJy]}} \\
\midrule
Quantity & RQG (Fiducial) & RQG (High-Z) & FRA-1 & FRA-2 \\
\midrule
F090W  & \multicolumn{2}{c}{$-3.7\pm7.7$}  & 0.6 & 0.3  \\
F115W  & \multicolumn{2}{c}{$45.2\pm7.9$}  &  61.6 &  13.9 \\
F150W & \multicolumn{2}{c}{$75.7\pm6.7$}  &  113.1 &  20.7 \\
F200W  & \multicolumn{2}{c}{$108.3\pm5.6$}  &  130.7 &  23.6 \\
F277W  & \multicolumn{2}{c}{$186.0\pm9.3$}  &  233.6 &  38.1 \\
F356W  & \multicolumn{2}{c}{$469.6\pm23.5$}  &  685.5 &  100.0 \\
F410M  & \multicolumn{2}{c}{$522.5\pm26.1$}  &  746.5 &  121.1 \\
F444W  & \multicolumn{2}{c}{$527.6\pm26.4$}  &  761.4 &  117.4 \\
F770W  & \multicolumn{2}{c}{$673.7\pm94.8$}  &  1061.8 &  155.7 \\
F1800W & \multicolumn{2}{c}{$377\pm1043$} &  1529.9 &  187.2 \\

\midrule
\multicolumn{5}{c}{\textbf{Physical Properties}} \\
\midrule

$\log_{10}(\mathrm{M_{\ast}/\mathrm{M_{\odot}}})$ & $10.23^{+0.04}_{-0.04}$ & $10.19^{+0.04}_{-0.04}$ & 10.31 & 9.46 \\
$\log_{10}(\mathrm{Z/\mathrm{Z_{\odot}}})$ & $-0.94^{+0.05}_{-0.04}$ & $0.07^{+0.08}_{-0.11}$  & 0.14 & -0.21 \\
$t_{50} \ / \ \rm{Gyr}$ & $0.20^{+0.07}_{-0.02}$ & $0.16^{+0.03}_{-0.02}$  & 0.30 & 0.31 \\
$\mathrm{A}_{\mathrm{V}}$ & $0.31^{+0.08}_{-0.08}$ & $0.25^{+0.09}_{-0.07}$  & 0.02 & 0.42 \\
$\mathrm{SFR_{100}} \ / \ \mathrm{M}_{\odot}\mathrm{yr}^{-1}$ & $0.84^{+20.16}_{-0.78}$ & $48.89^{+21.12}_{-13.04}$  & 1.44 & 0.51 \\
$\mathrm{SFR_{50}} \ / \ \mathrm{M}_{\odot}\mathrm{yr}^{-1}$ & $0.83^{+11.11}_{-0.76}$ & $2.13^{+5.54}_{-1.92}$  & 1.20 & 0.67 \\
$\mathrm{SFR_{10}} \ / \ \mathrm{M}_{\odot}\mathrm{yr}^{-1}$ & $0.64^{+0.83}_{-0.60}$ & $1.08^{+1.55}_{-0.98}$  & 0 & 0.91\\
\bottomrule
\end{tabular}

\caption{\emph{Top}: The observed photometry and associated $1\sigma$ uncertainties of RQG, alongside synthetic values for the \flares\ analogues as observed at $z_{\rm{RQG}}$. \emph{Bottom}: The physical properties of the same galaxies. In descending order, the stellar mass, stellar metallicity, age, V-band dust attenuation and star formation rates averaged over 100, 50 and $10 \ \rm{Myr}$ timescales. For RQG, we include the values inferred by both the fiducial and high-Z \prospector\ runs whereas the analogue values are derived directly from the particle properties at $z=7$. All observational results are taken from tables 1 and 2 in \protect\cite{Weibel_2025}.}

\label{tab:properties}
\end{table*}

\subsection{Observations}

RQG was identified in the UltraDeep Survey field from imaging taken as part of the PRIMER survey \citep[PI: Dunlop; See][]{Donnan_2024}. The photometric fluxes and uncertainties characterising the SED are listed by \citetalias{Weibel_2025} and reiterated here in Table~\ref{tab:properties}. These include eight NIRCam filters spanning $0.9 - 4.4 \ \mu\rm{m}$ in the observer-frame and two MIRI \citep{Wright_2023} bands centred on $7.7$ and $18.0 \ \mu\rm{m}$. Fluxes were measured in circular apertures of diameter $0.32$ and $0.5 \ \rm{arcsec}$ respectively, which are then scaled to total using a Kron aperture \citep{Kron_1980} and the encircled energy of the point spread function. We assume these corrections are accurate and treat the measurements as directly comparable to the total fluxes derived from simulated galaxies. Spectroscopic follow-up was performed as part of the RUBIES programme, which is introduced by \cite{de-Graaff_2025} alongside a summary of the data reduction. The PRISM/CLEAR configuration of the NIRSpec multi-shutter array was used to measure a spectrum from $0.6-5.3 \ \mu\rm{m}$ with $\rm{R} \sim100$ resolution. While this spectrum contains no well detected line features, the strong constraints on both the Lyman and Balmer breaks allow us to confidently assume that $z_{\rm{RQG}}=7.29$ is the true redshift. Where necessary, we use the spectrum provided by the DAWN JWST Archive \citep{brammer_2023, Heintz_2024}.

Two sets of physical properties were inferred by \citetalias{Weibel_2025} from the combined photometry and spectrum using the SED fitting code \prospector\ \citep{Johnson_2021}. Full details of both the `fiducial' and `high-Z' modelling approaches are provided therein and we reiterate the results in Table~\ref{tab:properties}. Both runs are identical bar the metallicity prior, which for the fiducial run is $\log_{10}(Z /Z_{\odot}) \in [-1, 0.19]$ and high-Z is $\log_{10}(Z /Z_{\odot}) > -0.5$. The latter is an attempt to marginalise over the absence of $\alpha$-enhancement in the FSPS stellar population synthesis (SPS) models \citep{Conroy_2010} used by \prospector\ during fitting, which could otherwise lead to the metallicity being underestimated. The measured $\chi^{2}$ is consistent between both runs, with the fiducial returning $\chi^{2}_{\mathrm{fiducial}} \approx 535$ and high-Z returning $\chi^{2}_{\mathrm{high-Z}} \approx 539$. This suggests an underlying bi-modality, where the high metallicity solution is only returned when the low metallicity solution is ruled out. As such, we consider both as potential solutions when validating \flares\ analogues. We analyse RQG in detail to facilitate a robust comparison, but it should be noted that all observational results are taken from \citetalias{Weibel_2025} and not original to this work.

\subsection{The First Light And Reionisation Epoch
Simulations}\label{sec:flares}

We search for theoretical analogues of RQG in the \flares\ simulations, which are described in detail by \citet{FLARES-I} and \citet{FLARES-II}. \flares\ is a suite of hydrodynamical zooms, based on a large $(3.2\ {\rm cGpc}^3)$ dark matter only simulation \citep{CEAGLE} that contains a number of rare environments. Forty spherical regions of radius $14/h\ {\rm cMpc}$ were selected from this box and re-simulated using the AGNdT9 variant of the \eagle\ model \citep{crain_eagle_2015, schaye_eagle_2015}, which best recovers the properties of group and cluster gas. At $z \sim 4.7$, these regions span overdensities $\delta + 1 = -1 \to 1$, with over representation towards the extremes to capture the most massive systems at early times. Global distribution functions are calculated after weighting each region to account for this preferential sampling \citep[See][section 2.4 and table A1]{FLARES-I}. Dark matter and initial gas particle masses of $9.7\times10^{6} \ \rm{M}_{\odot}$ and $1.8\times10^{6} \ \rm{M}_{\odot}$ respectively, coupled with a $2.66 \ \rm{ckpc}$ gravitational softening length, are more than sufficient to resolve massive galaxies. This approach makes \flares\ the ideal dataset in which to search for unique objects like RQG that may only form within extreme environments.

We generate new synthetic spectra and photometry of \flares\ galaxies for this analysis. These are forward modelled from particle data using \synthesizer\ \citep{Lovell_2025, Roper_2025}, accounting for the physical and dynamical properties of stars and gas. Models of AGN emission are still highly uncertain \citep{Wilkins_2025}, so black hole particles are excluded. We do not expect AGN to contribute significantly to the combined luminosity at these redshifts, particularly given the highly centrally concentrated dust distributions in massive galaxies in \flares\ \citep{FLARES-IV, FLARES-IX}. The sample is limited to subhalos consisting of at least one hundred star particles, ensuring that they are well resolved. Observables are measured within a $30 \ \rm{pkpc}$ aperture that captures the majority of the flux, but will not include distant gravitationally bound particles that are unlikely to be segmented observationally. Stellar particles are assigned spectra using the \texttt{BPASS-v2.2.1} \citep{BPASS2.2.1} SPS models, which include binary systems and the young stars expected to be commonplace at high redshift. These models do not include $\alpha$-enhanced spectra. We assume the stellar initial mass function of  \cite{ChabrierIMF}, ranging from $0.1-300 \ \rm{M_{\odot}}$. Reprocessing due to nebular gas is added post facto using the photoionisation code  \texttt{Cloudy} \citep{Cloudy17.02}, adopting the recipe of \cite{Wilkins_2020} which assumes an escape fraction $f_{\rm{esc}} =0$. Dust attenuation is modelled using the line of sight (LOS) model of \cite{FLARES-II}, in which stars are obscured by intervening gas particles assuming a dust optical depth derived from their metal density \citep{Vijayan_2019}. Additional attenuation is applied to stars formed in the most recent $10 \ \rm{Myr}$, to account for young stars still embedded within their dense birth clouds \citep{Charlot_2000}. Observer-frame photometry is finally derived from the appropriate NIRCam transmission curves and spectra resampled onto the NIRSpec wavelength grid using the \texttt{SpectRes} package \citep{Carnall_2017}. Both adopt the \cite{Inoue_2014} treatment for intergalactic medium (IGM) absorption.

\flares\ results are output at integer redshift snapshots from $5<z<15$, making $z=7$ the closest to $z_{\rm{RQG}}$ in both redshift and cosmic time. Absorption and emission line features can cause the broadband fluxes measured from otherwise flat spectra to vary drastically over only small increments in redshift and could introduce bias when comparing \flares\ galaxies to RQG. To minimise this effect, we generate synthetic observables at both the fiducial redshift and as they would appear at $z_{\rm{RQG}}$. The latter is used exclusively in Section~\ref{sec:analogues} to identify candidate analogues and compare their observable properties, while the former is used when deriving physical properties in the observer-frame (\S\ref{sec:alpha}). This ensures consistency in both cases, but is only viable when the difference in time is small. In this case the shift is equivalent to only $\sim 40 \ \rm{Myr}$, but should still be considered when comparing physical properties.

%% file: Sections/3_analogues.tex
\section{Identifying Analogues} \label{sec:analogues}

\begin{figure}
 	\includegraphics[width=\columnwidth]{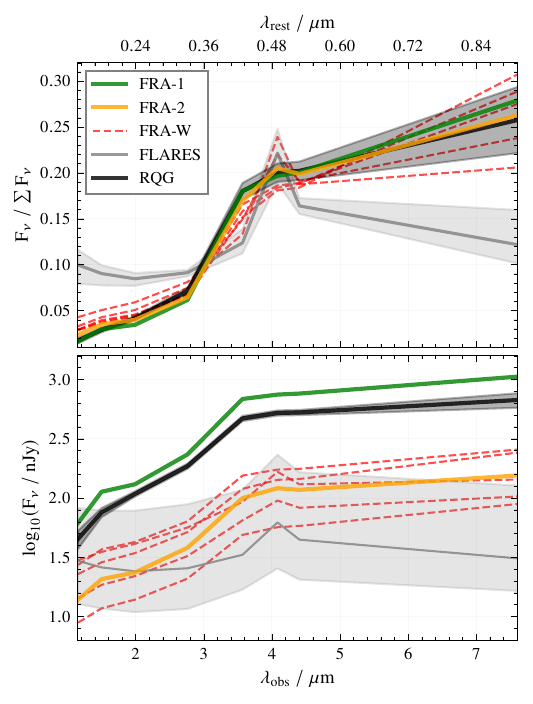}
	\caption{\emph{Top}: The SEDs of RQG (black) and the potential \flares\ analogues as observed at $z_{\rm{RQG}}$ after normalising by the total flux. The best matches FRA-1 and FRA-2 are indicated by green and orange lines respectively. Dashed red lines show the weaker matches. The grey line is a median \flares\ SED, constructed by taking the median value in each filter individually after normalisation. Shaded regions indicate $1\sigma$ confidence limits. Straight lines join the wavelength dependent measurements of each galaxy for readability, but are not physically representative. \emph{Bottom}: The absolute SEDs of the same galaxies. 
	\label{fig:seds_comp}}
\end{figure}

\subsection{Matching SEDs}\label{sec:matching}

There are many means of identifying analogues of observed galaxies in simulations. Studies of the Milky Way often select objects based on their physical properties, to find galaxies that are identical at the time of observation and are therefore likely to have evolved in the same way \citep{Calore_2015, Pillepich_2024}. While these quantities are accurate up to the resolution limit in \flares, for RQG they are inferred through SED fitting which is a process fraught with systematic effects \citep{Narayanan_2024, Harvey_2025}. These are not accounted for in the quoted uncertainties, so could introduce bias if used for matching. In contrast, the photometry itself can be measured reasonably accurately by \emph{JWST} and the uncertainties are comparatively well understood \citep{Dicken_2024, Merlin_2024}. Using these quantities alone should facilitate a robust matching. We therefore focus on the SED itself, characterised by broadband fluxes, which should encode more information than the physical properties alone \citep{Conroy_2013, Iyer_2025}. These fluxes are less sensitive to noise and cover a broader wavelength range than the spectrum, which we do not consider when matching. 

The bottom panel of Figure~\ref{fig:seds_comp} compares the SED of RQG to the median measured from all \flares\ galaxies with at least one hundred star particles. This highlights its exceptional brightness. Given its high inferred stellar mass, it is not surprising that RQG is a factor $1.3 \ \rm{dex}$ brighter than average at the longest wavelengths. This by no means indicates that galaxies as extreme as RQG do not exist in \flares, but they are likely to be extremely rare. Therefore, to expand the potential sample size and guard against coincidental matches, we focus our search purely on the shape of the SED. While the best analogues should have comparable brightness, it is not unlikely that other massive quiescent galaxies could be fainter and evolve under similar mechanisms. The absolute value of a non detection is primarily dependent on the local noise level of the imaging and is also sensitive to the source extraction approach. We would therefore require full image based forward modelling to produce comparable values, which would still hold very little physical information about the source. While these can be treated as upper limits, we choose to omit F090W and F1800W from the matching, as including them without this modelling could introduce bias. The remaining eight fluxes are normalised by their sum to produce an SED that preserves shape information in each filter. The dissimilarity between RQG and each \flares\ galaxy is then quantified by a score $D$. To account for these observational uncertainties, we perform 10000 matching iterations where the RQG flux in each band $b$ is resampled from a Gaussian with width defined by the quoted error. The total dissimilarity score of each \flares\ galaxy is then simply \begin{equation}
    D = \sum_{i}^{10000}{\sum_{b}{\frac{f_{b,i}-F_{b}}{f_{b,i}}}} \,,
\label{eq:match}\end{equation} where $F$ denotes the normalised flux of the galaxy and $f$ is the normalised and resampled flux of RQG. We record the closest match in each iteration, defined as the galaxy with the lowest value of the inner sum of equation \ref{eq:match}.

A potential analogue must be the closest match in at least one iteration and we find only seven such galaxies in our sample. We refer to these as \flares-RQG Analogues (FRA) herein. The quality of a match is then determined by the dissimilarity score, with lower values indicating a better match. The top panel of Figure~\ref{fig:seds_comp} shows the total flux normalised SED of RQG used to perform this matching, alongside the \flares\ median. The two clear analogue candidates are FRA-1 and FRA-2, which are the best match in $55\%$ and $42\%$ of resampling iterations respectively. The former is the best overall match, with a dissimilarity score $D\sim113$, but this is only marginally lower than FRA-2 with $D\sim121$. The SEDs trace RQG across all wavelengths, deviating only slightly in F200W-F277W colour and demonstrating that their matching is not coincidental. The flux of both galaxies in F090W and F1800W is consistent with $1\sigma$ and $2\sigma$ RQG upper limits respectively. Given these sources are the best matches in $>95\%$ of the resampling iterations, we consider them the most promising analogues and list their properties in Table~\ref{tab:properties}. The remaining five candidates are grouped as `weak' analogues (FRA-W) in the following analysis, as they were each the best match in only $\sim1\%$ of resampling iterations and have far larger dissimilarity scores. These SEDs are not as comparable as FRA-1 and FRA-2 and are incapable of tracing RQG over any extended wavelength interval. Two of these galaxies display a bright F410M component, more akin to the \flares\ median. This can also be seen very subtly in FRA-2 and hints at stronger emission line features, associated with star-forming systems. We discuss the SPS model dependence of this process in Appendix~\ref{app:matching}. Matching to observables derived from FSPS, the model assumed by \citetalias{Weibel_2025}, recovers the same two galaxies as the best matches. Our interpretation of the underlying physics in these galaxies would therefore be be unaffected by adopting this model.

\subsection{Observable Properties}\label{sec:observables}

\begin{figure}
 	\includegraphics[width=\columnwidth]{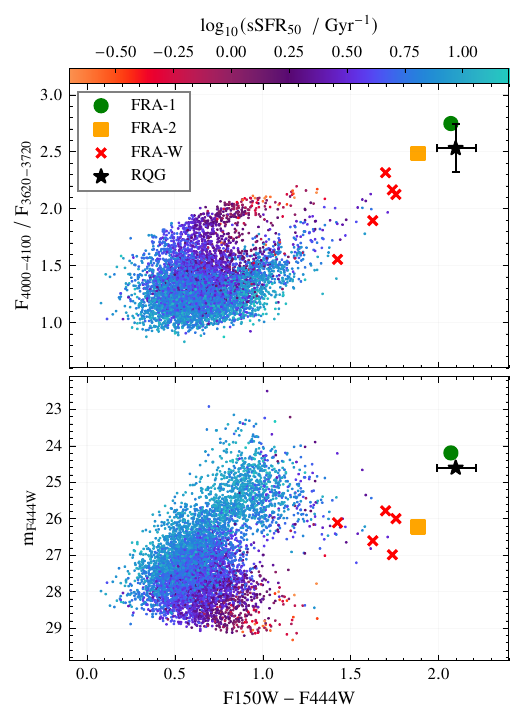}
	\caption{\emph{Top}: The Balmer break strength measured using the definition of \protect\cite{Wang_2024} as a function of F150W-F444W colour, quantifying the redness of the SED. RQG is represented by a black star, its closest analogues by green and yellow points and the weak candidates by red crosses. Other \flares\ galaxies are coloured by their sSFR, averaged over the most recent $50 \ \rm{Myr}$. \emph{Bottom}: AB-magnitude in F444W as a function of redness for the same galaxies. All observational quantities are measured at $z_{\rm{RQG}}$.}
    \label{fig:balmer}
\end{figure}

\begin{figure*}
        \centering
 	\includegraphics[width=\textwidth]{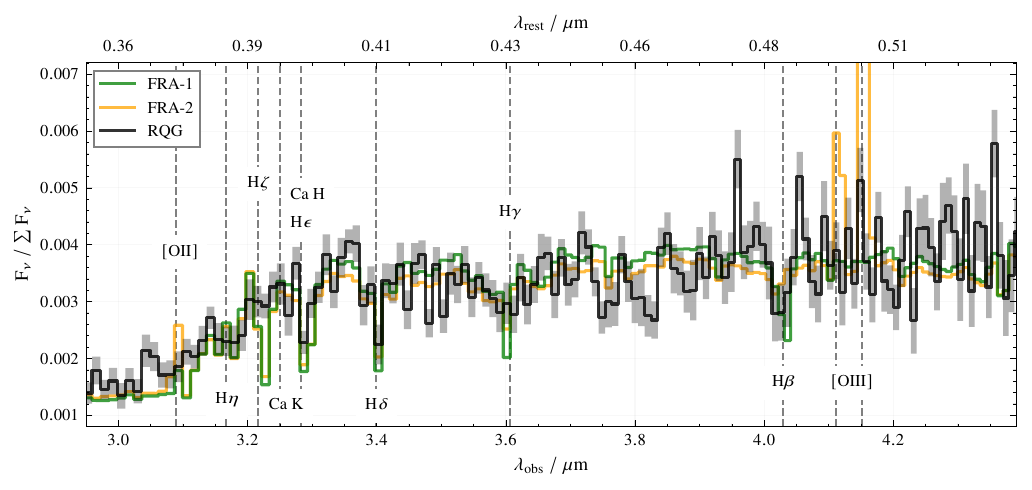}
	\caption{A portion of the NIRSpec/PRISM resolution spectrum of RQG (black) and the best two \flares\ matches (green and orange). These are as observed at $z_{\rm{RQG}}$ and normalised by the total flux. Weak matches are omitted for readability. The shaded region indicates the $1\sigma$ confidence limits on the RQG spectrum and the locations of key absorption and emission features are indicated by dashed lines.
	\label{fig:lines}}
\end{figure*}

We can now investigate the properties of these galaxies in more detail, to verify their analogous nature. Given the aforementioned uncertainties in 
SED fitting, we first investigate the agreement between RQG and its analogues using direct observables. The bottom panel of Figure~\ref{fig:seds_comp} also shows the absolute SEDs of these objects. Perhaps surprisingly, RQG is by no means the brightest and is outshone by FRA-1 at all wavelengths. This $\sim 0.2 \ \rm{dex}$ difference could be attributed to additional stellar mass formed since $z_{\rm{RQG}}$, but an average SFR of $1.2 \ \mathrm{M_{\odot} yr^{-1}}$ would only increase the mass by $\sim 0.2\%$. RQG may be bright, but \flares\ is able to produce brighter galaxies, which may well be passive, at even earlier cosmic times. The remaining analogues are fainter by up to $1\ \rm{dex}$, making them more consistent with the \flares\ median. While unlikely to be as analogous or extreme as FRA-1, they may still be subject to the same mechanisms, so we do not discard candidates at this stage.

The bottom panel of Figure~\ref{fig:balmer} shows the apparent AB-magnitude measured in F444W against the F150W-F444W colour used by \cite{de-Graaff_2025} to quantify the redness of the SED. There are many galaxies over $1 \ \rm{mag}$ brighter than RQG in this long wavelength filter, further highlighting that it is not extreme in luminosity space. However, these galaxies are all far bluer and have high sSFRs, indicating that their brightness is driven by more luminous young stars. Given the inferred recent SFR of RQG, we expect it to instead be driven by a large population of fainter intermediate and old stars and this difference hints a potential offset in age. What clearly makes RQG unique is its redness, which is $0.5 \ \rm{dex}$ higher than any non-analogue galaxy. The best matches are the two reddest galaxies in \flares, with FRA-1 overlapping almost exactly. Unsurprisingly, the weak matches are also weaker in their redness, but all fall within the top percentile. Given the rest-frame wavelength probed, this again implies an older than average stellar population, which would require these galaxies to begin meaningful star formation far earlier and in the case of RQG and FRA-1 more rapidly. This is consistent with the $\sim300 \ \rm{Myr}$ ages of FRA-1 and FRA-2 listed in Table~\ref{tab:properties}, defined by the time since $50\%$ of the current stellar mass formed. 

The top panel of Figure~\ref{fig:balmer} compares the same redness criterion to the strength of the Balmer break, quantified using the definition of \cite{Wang_2024} which avoids strong spectral features. RQG again occupies a unique region of the parameter space, alongside its main analogues which share the strongest breaks in \flares. This break is commonly used as an indicator of age \citep{Bruzual_1983, Poggianti_1997}, but also as an indicator of quiescence. Indeed \flares\ galaxies with the strongest breaks tend to have lower sSFRs, although this may start to break down in redder galaxies. Furthermore, \cite{Wilkins_2023b} showed that instantaneous and declining star formation histories, both consistent with RQG, produced stronger breaks than increasing or constant histories under the same conditions. We can therefore interpret the top panel as we did the bottom and say that these galaxies formed early and rapidly.

Figure~\ref{fig:lines} shows the NIRSpec spectrum of RQG after normalising by the total flux and zooming on the region with the clearest features. The Balmer absorption lines are consistent with noise individually, but are more confidently identified by eye when considered together. Those most clearly identifiable, such as $\rm{H}\beta$ and $\rm{H}\delta$, are produced most intensely by the atmospheres of long lived A-type stars \citep{Poggianti_1997}. Combined, these features are often associated with the post-starburst scenario, in which galaxies undergo a major burst of star formation that ceases rapidly within $1 \ \rm{Gyr}$ of observation \citep{Couch_1987, Zabludoff_1996}. This burst likely happened early, as the absence of emission lines from ionised gas suggests very few young stars remain. Spectra of the best \flares\ analogues are also plotted, after resampling with \texttt{SpectRes} to match NIRSpec resolution. Both show the same absorption features, including far stronger $\rm{H}\gamma$ and $\rm{H}\zeta$, which imply a comparable burst dominated history. \citetalias{Weibel_2025} report no well detected emission lines and the same can be said of FRA-1 over the full wavelength range. However, FRA-2 shows a clear doublet of [O\textsc{iii}]$\lambda4959, \ \lambda5007$ and [O\textsc{ii}] emission, revealing the presence of an ionising source. Given that AGN are not included in the forward modelling, this must be a young stellar population. While this could be a significant deviation from RQG, we are likely observing this galaxy $> 100 \ \rm{Myr}$ later in its evolution when we consider the differences in age. It is not unfeasible that it evolved in the same way prior to this recent increase in activity, and previous observations suggest this population is insignificant. As such, we opt not to discard FRA-2 as a potential analogue. 

In general, we find the observables of RQG to be unusual relative to the general \flares\ population, but the identified analogues are consistent and in some cases even more extreme. As expected, they are characterised by their red colours and strong Balmer breaks, although high luminosity and a lack of any emission lines are not a requirement to achieve this.  \flares\ is clearly able to produce galaxies that look like RQG and likely share similar histories. 
 
\subsection{Physical Properties}\label{sec:physical}

\begin{figure}
 	\includegraphics[width=\columnwidth]{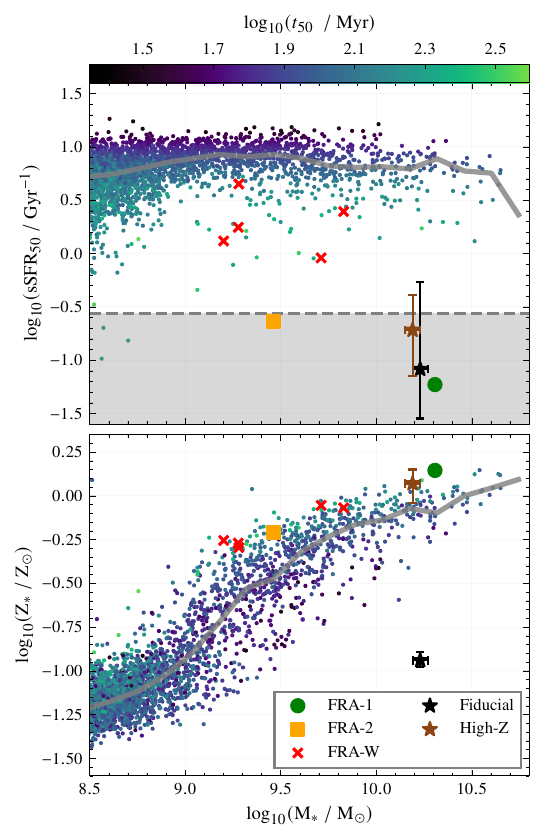}
	\caption{\emph{Top}: The sSFR-stellar mass relation as predicted by \flares. The position of RQG is shown for both the fiducial (black) and high-Z (brown) \prospector\ models. The two best matches are denoted by green and orange points and the weak matches by red crosses. The grey box indicates the selection region for quiescent galaxies at $z_{\rm{RQG}}$, using the time dependent criterion defined by \protect\cite{Pacifici_2016}. \emph{Bottom}: The mass-metallicity relation of the same objects. \flares\ metallicities are averaged over all stellar particles, after weighting by their current masses. Grey lines indicate the median value of each quantity as a function of stellar mass.
	\label{fig:ssfr_z}}
\end{figure}

We can now determine whether these analogues are also massive and quiescent. Figure~\ref{fig:ssfr_z} demonstrates where RQG and its analogues fall on two key relations measured within the \flares\ model, with the fiducial and high-Z observational fits indicated by black and brown points respectively. Throughout this work, we define stellar mass as the current mass in stars at the time of observation, consistent with SED fitting inferred measurements. The top panel shows the sSFR-stellar mass relation, which forms a roughly mass invariant main sequence (MS) at $\mathrm{sSFR_{50}} \sim 0.8 \ \mathrm{Gyr}^{-1}$, that is consistent with studies at both low and comparable redshift \citep{FLARES-VII, Nakajima_2023}. The grey shaded box indicates the selection region for quiescent galaxies, employing the age of the Universe dependent criterion 
\begin{equation}
    \mathrm{sSFR} < \frac{0.2}{t_{\mathrm{uni}}(z)} \,, \label{eq:ssfr_cut}
\end{equation} defined by \cite{Pacifici_2016} and regularly adopted by observational studies \citep{Carnall_2023,  Alberts_2024, Russell_2024}. The value plotted assumes $z=z_{\rm{RQG}}$, but using the fiducial snapshot redshift has no noticeable effect on the extent of the region or the galaxies selected. It should be noted that selections in this space are heavily dependent on the aperture and SFR averaging timescale used \citep{Donnari_2019, Donnari_2020}. For consistency with the photometry, we use a $30 \ \rm{pkpc}$ aperture for both quantities and follow \cite{FLARES-VIII} in adopting a $50 \ \rm{Myr}$ timescale to capture both recent and intermediate age star formation.

There are only two massive quiescent galaxies at this redshift, FRA-1 and FRA-2. Again, FRA-1 appears to be an excellent analogue, with $2\sigma$ and $3\sigma$ agreement in mass and $1\sigma$ and $2\sigma$ agreement in $\mathrm{sSFR_{50}}$ for the fiducial and high-Z runs respectively. It occupies a unique region of parameter space, being both the most massive and least star forming quiescent galaxy in our sample by $0.8$ and $0.3 \ \rm{dex}$ margins, respectively. While this again highlights RQG as an extreme case, it by no means matches the most massive galaxies. Despite FRA-1 being within the top percentile, there are galaxies with up to $\sim0.5 \ \rm{dex}$ more stellar mass. It is clearly not a challenge for \flares\ to produce purely massive galaxies this early in cosmic time. The remaining galaxies are coloured by their age, quantified by the time since forming $50\%$ of their stellar mass. Massive galaxies are systematically younger than RQG and quenching has not yet had the opportunity to take effect. This again points to formation time as a key driver of massive quiescent galaxies in this epoch. While FRA-2 is both massive and quiescent, the stellar mass is $0.85 \ \rm{dex}$ lower, so cannot be considered an extreme example. Interestingly, the weak analogues have varying masses, but all fall below the MS while remaining at least $\sim 0.5 \ \rm{dex}$ above the selection region. This could imply that they are progenitors of RQG like galaxies that may become quiescent in the near future.

The bottom panel shows the position of the same galaxies on the stellar mass-metallicity relation, which has also been shown to agree well with observations \citep{Nakajima_2023}. \flares\ metallicities are averaged over all star particles belonging to the subhalo, after weighting by their mass. While SED fitting results are luminosity weighted by nature, we do not expect large differences between the two schemes \citep{Guidi_2015, Fraser-McKelvie_2021}, particularly in quiescent galaxies that lack bright young stars. All analogues lie on the MS, but are preferentially enriched relative to the general population at each mass. This region is occupied by the oldest galaxies, which have had more time to disseminate the metals synthesised in the early stages of star formation into their environment and new stars. However, some weak analogues are $150 \ \rm{Myr}$ younger than FRA-1 and FRA-2 and the inferred age of RQG is only $\sim200 \ \rm{Myr}$, so an additional less age-dependent mechanism must be partly responsible. If these galaxies have already consumed the nearby gas supply, they would appear preferentially enriched, as the metallicity is not regulated by stars forming from pristine IGM gas. This lack of inflowing gas would also cause the systematically low sSFRs, producing the well studied fundamental metallicity relation \citep{Mannucci_2010}. The gas properties are investigated further in Section~\ref{sec:quiescense}. Perhaps most intriguing is the $1 \ \rm{dex}$ difference in the metallicities inferred by the different \prospector\ models. While the fiducial run is at least $\sim 0.7 \ \rm{dex}$ lower than the \flares\ expectation given its mass, the high-Z run falls on the same line as its analogues, consistent with the metallicity of FRA-1. \flares\ clearly favours a super solar metallicity in RQG-like galaxies, which could mean the fiducial run significantly underestimates the value and its associated uncertainty. \citetalias{Weibel_2025} suggest that this discrepancy could be due to significant $\alpha$-enhancement, which we evaluate in Section~\ref{sec:alpha}.

We have verified that \flares\ galaxies most closely mimicking the observable properties of RQG are in fact massive quiescent galaxies. The model is clearly able to produce comparable systems early in the Universe's history, although there may still be some tension in their chemical enrichment under the assumption of a fiducial \prospector\ result. Despite sharing many similar properties, galaxies only partially resembling RQG are not quiescent and therefore discarded.

\subsection{Star Formation History}

\begin{figure}
 	\includegraphics[width=\columnwidth]{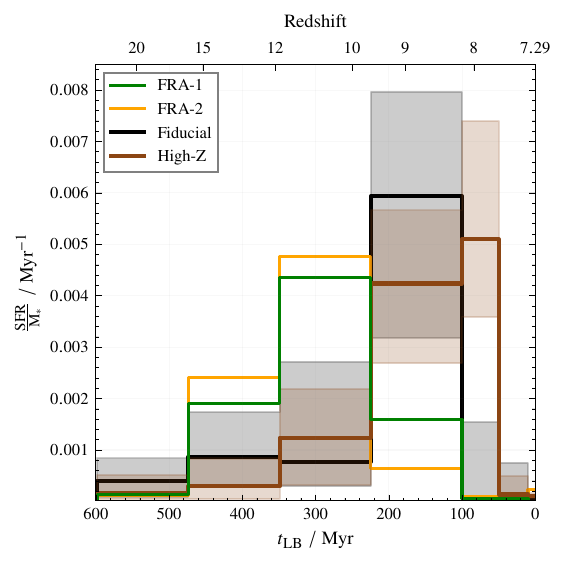}
	\caption{The star formation history of RQG as a function of lookback time inferred by the fiducial (black) and high-Z (brown) \prospector\ models. Shaded regions indicate the $1\sigma$ confidence limits reported by \protect{\citetalias{Weibel_2025}} and the two \flares\ analogues are shown by green and orange lines. Each history is normalised such that it integrates to one, representing the relative rate in each time interval. 
	\label{fig:normed_sfh}}
\end{figure}

The final verification step to ensure we have selected strong RQG analogues is comparing SFHs. The SFHs of quiescent galaxies have been shown to vary greatly across models, so should provide a useful constraint \citep{Lagos_2024b}. Differences in stellar mass have already been highlighted, so we again focus on the shape to establish whether they assemble in the same manner. \citetalias{Weibel_2025} use a redshift dependent binning, but for ease of comparison the same bins are used for the \flares\ analogues. These consist of $\sim 125 \ \rm{Myr}$ width bins up to $100 \ \rm{Myr}$ before observation, with the remaining time divided into bins of $50$, $40$ and $10 \ \rm{Myr}$. \flares\ histories are measured by summing surviving stars over all progenitor branches, mimicking the nature of the observations. The normalised histories are shown in Figure~\ref{fig:normed_sfh}, where the shaded regions indicate the $1\sigma$ confidence limits obtained by \citetalias{Weibel_2025}. These emphasise the degree of uncertainty, as the dominant peak can produce anywhere between $40 - 99\%$ of the RQG stellar mass. Taking the lower value, the timescale over which the stellar mass forms becomes far more standard, ramping up steadily over $\sim500 \ \rm{Myr}$. Of course the opposite could also be true in which case RQG would be the most burst dominated massive galaxy ever identified. The end of star formation is thankfully more certain, with the quenching mechanisms taking hold abruptly in both scenarios and reducing the peak to almost zero over $100 \ \rm{Myr}$ at most. The most obvious difference between the fiducial and high-Z results is the nature of the burst, with the latter being weaker but extended by $50 \ \rm{Myr}$ in order to form a comparable stellar mass. 

These uncertainties propagate to the comparison of analogues, but we attempt to use the SFHs to make a qualitative comparison regardless. FRA-1 and FRA-2 are both older than RQG, so prominent features in their SFHs should appear $\sim 100 \ \rm{Myr}$ earlier. Both galaxies are dominated by a single burst which forms $55\%$ and $60\%$ of the stellar mass respectively, consistent with both RQG runs. The absolute masses formed in the bursts of RQG and FRA-1 are therefore comparable when assuming the median value. The start and end of star formation appear marginally more gradual in both analogues, but given the coarse binning it is hard to differentiate between this and a more extended burst. In FRA-1, $20\%$ of the stellar mass is formed on either side of the burst and FRA-2 forms $25\%$ before, but both are within the RQG uncertainties. Interestingly, FRA-2 forms $\lesssim10\%$ of its stellar mass at the end of the burst. This  suggests that the quenching mechanisms may take hold more quickly, despite it being less massive. It is only a small increase in SFR over the most recent $50 \ \rm{Myr}$ that make it appear less quiescent in Figure~\ref{fig:ssfr_z} and this activity is also responsible for the previously identified emission lines.

Both FRA-1 and FRA-2 have demonstrated good agreement with the observable properties of RQG and are identified as massive and quiescent. In addition, their SFHs are each dominated by a rapidly quenched burst of comparable intensity and duration. Despite FRA-2 being less massive and beginning to form some stars over the most recent $10 \ \rm{Myr}$, these galaxies clearly formed in a very similar way to RQG. We therefore consider both strong analogues. This confirms that \flares\ can produce RQG-like galaxies, which can now be used to examine the underlying physics.

%% file: Sections/4_physics.tex
\section{Discussion}\label{sec:physics}

\subsection{Forming an RQG-like Galaxy}\label{sec:quiescense}

\begin{figure}
 	\includegraphics[width=\columnwidth]{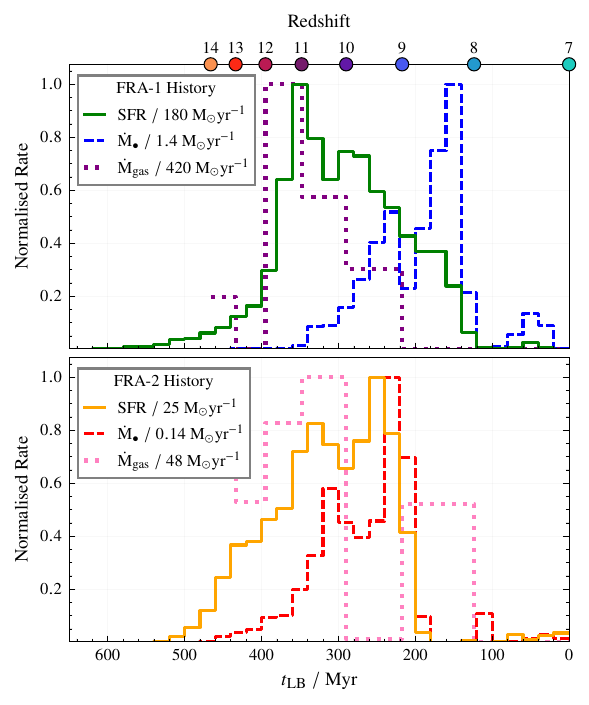}
	\caption{The normalised star formation (solid), AGN accretion (dashed) and gas accretion (dotted) histories of FRA-1 and FRA-2, as measured over the entire subhalo. Normalisation factors are indicated in the legends. Star formation and AGN accretion histories are binned in look back time intervals of $20 \ \rm{Myr}$, while gas accretion is plotted at snapshot resolution. Redshift ticks on the top axis are replaced by coloured points matching those of Figure~\ref{fig:gas_frac}, to ease comparison.
	\label{fig:accretion}}
\end{figure}

\begin{figure}
 	\includegraphics[width=\columnwidth]{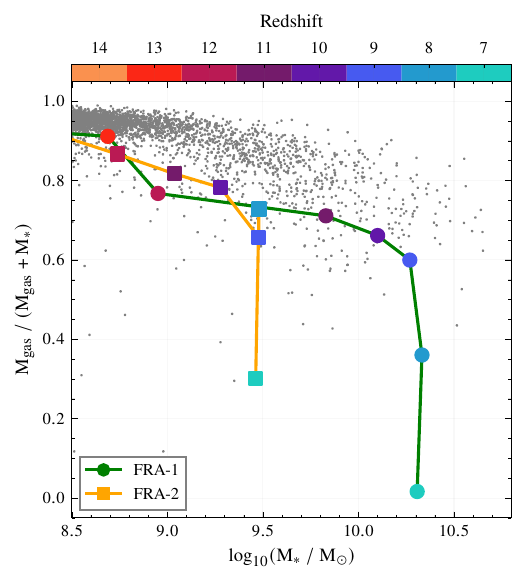}
	\caption{The fraction of baryonic mass in gas as a function of stellar mass, measured over the entire subhalo. Grey points show the position of all \flares\ galaxies at $z=7$, and coloured tracks show the evolution of FRA-1 (green) and FRA-2 (orange) in this space using the main progenitor at each redshift.
	\label{fig:gas_frac}}
\end{figure}

We also plot the normalised SFHs of FRA-1 and FRA-2 in Figure \ref{fig:accretion}, now at $20 \ \rm{Myr}$ time resolution as we are no longer limited by the observations. We also show the gas accretion history, calculated as the difference in gas mass between progenitors at each snapshot redshift, after accounting for gas converted into stars. The SFR of FRA-1 increases gradually for $200 \ \rm{Myr}$, before more than doubling in only $40 \ \rm{Myr}$ at $z=12$ to reach its $180 \ \rm{M_{\odot}yr^{-1}}$ maximum. By $z=11$ it has become the most star forming galaxy in \flares\ and accrued the second most stellar mass. Despite undergoing such an intense burst, the gas mass increases by $0.75 \ \rm{dex}$ over its duration. This reveals a large inflow from the surrounding IGM, with an average gas accretion rate $\dot{\mathrm{M}}_{\bullet} = 420 \ \mathrm{M_{\odot}yr^{-1}}$ fuelling the burst. For comparison, the most massive \flares\ galaxy at this time undergoes a later but stronger burst, while also increasing its gas mass by $>1 \ \rm{dex}$. Those among the most massive by $z=7$ are younger, form their mass more steadily and do not suddenly increase their gas mass. While not a requirement for producing massive galaxies in general, early formation requires a sudden burst, facilitated by a large volume of cool and pristine external gas. In FRA-2 the SFR and gas mass increase more gradually. Gas is accreted at a maximum rate of only $\dot{\mathrm{M}}_{\bullet} = 48 \ \mathrm{M_{\odot}yr^{-1}}$, indicating that it does not have access to this same local resource and as such cannot reach a comparable stellar mass.

\cite{FLARES-VIII} investigated the causes of passivity in a broad population of \flares\ galaxies and identified black hole feedback as the primary driver, consistent with \emph{JWST} observations at intermediate redshift \citep{Kokorev_2024, Ito_2025}. To investigate whether this is also responsible in RQG, we also plot the total black hole accretion rates in Figure~\ref{fig:accretion}, at $20 \ \rm{Myr}$ time resolution. Both galaxies host only the original black hole seeded when the subhalo mass exceeded $10^{10} \ h^{-1} \mathrm{M_{\odot}}$, so this quantity is equivalent to the AGN accretion rate. 

Despite an abundance of gas, the SFR in FRA-1 decreases by $40\%$ over the same timescale that it increased, without any significant accretion onto the black hole. While the galaxy is not yet quenched by any extent, star formation is clearly being regulated by a mechanism other than AGN feedback at this time, as it is yet to begin significant accretion. A rapid and strong burst formed of pristine gas is likely to produce a high number density of massive stars and supernovae \citep{Abel_2002, Yoshida_2006}. In \flares, these are modelled at the subgrid level and produce winds that stochastically heat the surrounding ISM and expel star forming gas \citep{Dalla_2012}. These effects work in tandem to produce the decrease in SFR at $10<z<11$. However, by $z=10$ the gas mass has increased by $0.2 \ \rm{dex}$, demonstrating that stellar feedback is incapable of fully ejecting gas from the subhalo. While star formation may be disrupted for some time, the required gas is still available, meaning full and prolonged quenching cannot be achieved. We can also see that stellar feedback acts over short timescales, as the SFR increases again after only $40 \ \rm{Myr}$. Once massive stars and supernovae have expired, they cannot prevent the gas cooling and coalescing once more \citep{Iyer_2020}. While stellar feedback may be first to take effect and can somewhat regulate star formation, it is clearly not sufficient to fully quench massive galaxies, even after such an intense burst. These results are consistent with higher resolution simulations \citep{Su_2019, Bassini_2023} and expected in the \eagle\ subgrid model, given that stellar feedback has been shown to be highly inefficient \citep{crain_eagle_2015}. 

The AGN begins to accrete at the same time, reaching its first peak at $z\sim9$, before doubling in the following $80 \ \rm{Myr}$. These increases correlate with the accelerated decline of star formation at $z\sim10$ and its eventual cessation by $z=8$. This correlation highlights AGN feedback as the primary driver of quiescence. While the gas accretion rate also decreases during this time, by referring to Figure \ref{fig:gas_frac} we can see that gas still makes up $60\%$ of the baryonic mass at $z=9$. Therefore, while FRA-1 may be in the early stages of strangulation, it cannot yet be responsible for the end of star formation. The effect of AGN feedback is seen even more clearly in FRA-2, where accretion peaks only $20 \ \rm{Myr}$ after the peak of star formation. Feedback then reduces the star formation rate to zero in only $80 \ \rm{Myr}$, consistent with short timescales inferred by \cite{Sato_2024}. In fact, both SFR drops in FRA-2 coincide with peaks in AGN accretion, again implying that stellar feedback has little effect and that AGN feedback is the primary SFR regulator.

The cessation of star formation in both FRA-1 and FRA-2 correlates with a sudden drop in AGN activity, explaining the low AGN luminosities inferred from intermediate redshift quiescent galaxies \citep{Bugiani_2025} and the lack of X-ray detections \citep{Almaini_2025} in comparable systems. The black holes have grown significantly by this time, with mass ratios $\mathrm{M_{\bullet}/M_{\ast}} = 0.9\%$ and $0.4\%$ in FRA-1 and FRA-2 respectively. When star formation begins to ramp up again in both galaxies, we also see the AGN restarting its accretion. These AGN are already sufficiently massive to prevent timely rejuvenation and in the case of FRA-1 return the SFR to exactly zero. AGN are therefore not only the drivers of initial quenching, but they also act to maintain it for an extended period.

In FRA-2, these mechanisms appear to also regulate the initial formation stage. Its more steady and stunted SFH appear to be due to the AGN becoming relatively more active earlier in its lifetime. This could be due to a higher concentration of gas within the vicinity of the black hole. At $z=10$, over $50\%$ of the gas particles are within $1 \ \rm{kpc}$ of the AGN, compared to only $30\%$ in FRA-1. While the \eagle\ subgrid model does not directly link the strength of feedback to particle proximity, these events will occur more regularly if more gas is available for accretion. Angular momentum and the relative velocities of the gas and black hole also play a role.

To better understand why AGN feedback is so effective, we plot the relation between gas mass fraction and stellar mass in Figure~\ref{fig:gas_frac}, with both quantities measured over the entire subhalo. The general \flares\ population are shown at $z=7$, with tracks of the analogues' main progenitors at each redshift over-plotted. More massive galaxies tend to have lower gas fractions, as they are by nature more efficient at turning the same gas into stars. As discussed above, stellar feedback is ineffective at removing gas, causing the fraction to decrease by only $\sim 10\%$ in FRA-1 during its most intense star forming period. Once the AGN accretion rate reaches maximum, feedback reduces the fraction from $60$ to $<1\%$ in only $\sim 150 \ \rm{Myr}$. This makes FRA-1 the least gaseous galaxy at $z=7$, highly discrepant with those of comparable mass. This shows that unlike stellar, AGN feedback is able to fully expel gas from the subhalo, increasing the time required to rejuvenate and limiting future star formation if it is further removed from the gravitational well. \cite{Valentino_2025} tentatively measured a gas outflow rate $\dot{\mathrm{M}}_{\mathrm{out}}\sim 269 \ \rm{M_{\odot}yr^{-1}}$ in RQG, a factor of three greater than the rate in FRA-1. This was inferred from the blue shifting of Mg\textsc{II} absorption lines, so only includes gas expelled towards the observer. Considering the peak of activity in FRA-1 has long passed, these are still both extremely high outflow rates \citep{Leung_2019, Schreiber_2019}. The authors suggest that an undetected AGN may be responsible, which these results corroborate. Detection is likely to be challenging and tied to the current rate of star formation, as activity is limited while the gas fraction remains low.

While the removal of gas is clearly highly efficient, the galaxy is already fully quiescent by $z=8$, when $35\%$ of the mass is still in gas. In FRA-2, the gas fraction falls neither as quickly or to the same level and $30\%$ of the mass remains in gas by $z=7$. The gas fraction briefly increases after the peak in AGN activity, before dropping at a rate $\dot{\mathrm{M}}_{\mathrm{out}}\sim 55 \ \rm{M_{\odot}yr^{-1}}$. Comparable galaxies are able to form stars under these conditions, so both analogues already being quiescent highlights the heating of gas by AGN as equally important. It appears to take hold more quickly than gas expulsion and is therefore responsible for the appearance of the earliest massive quiescent galaxies. While expulsion also contributes to this, it primarily ensures that quenching persists for an extended period. Interestingly, these effects have an imprint on the levels of dust attenuation. The few gas particles remaining in FRA-1 are unable to effectively obscure the LOS and the resulting $\mathrm{A_{V}}$ is almost negligible, whereas FRA-2 falls only marginally lower than expected given its stellar mass \citep{Cullen_2018}. Applying a similar logic to RQG, which has reasonably high $A_\mathrm{V}$, we could infer that a gas fraction comparable to that of FRA-2 remains and that heating is still the current driver of its total quiescence. This is consistent with the high outflow rate, but uncertain without measurements of the rest-frame infrared emission to better constrain the dust properties \citep{Buat_2012, Hayward_2014, Salim_2020}. This is however complicated, as the nature of the outflow measurement from RQG implies that at least some fraction of the expelled dust and gas still obscures the LOS. Furthermore, we demonstrate in Section~\ref{sec:alpha} and Appendix~\ref{app:prospector} that dust attenuation may be systematically overestimated.

\subsection{Metallicity and $\alpha$-enhancement}\label{sec:alpha}

\begin{figure}
 	\includegraphics[width=\columnwidth]{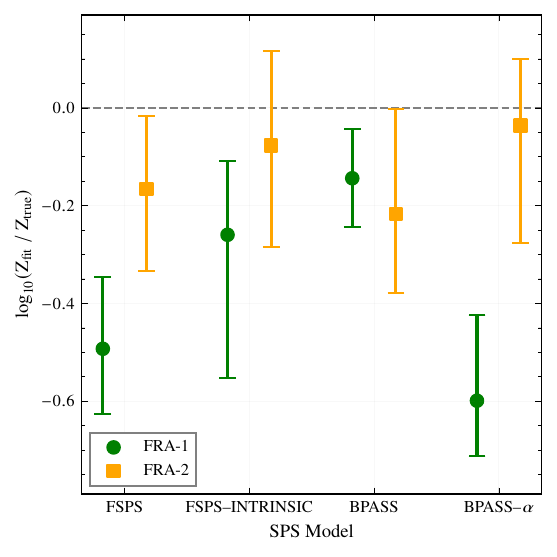}
	\caption{The ratio between \prospector\ inferred and true metallicity for the RQG analogues. These values are quoted for the four different models used to generate the spectra and photometry. Fitting was performed using the same FSPS model in each case.
	\label{fig:metal}}
\end{figure}

The level of chemical enrichment is one of the biggest questions surrounding the nature of RQG. The $1 \ \rm{dex}$ discrepancy in metallicities inferred by the fiducial and high-Z \prospector\ runs make drawing conclusions difficult and \citetalias{Weibel_2025} attribute this to high levels of $\alpha$-enhancement. Such an assumption is reasonable given the effect on stellar spectra \citep{Vazdekis_2015, Byrne_2023} and expectations from the literature at high-redshift \citep{Maio_2015, Ma_2017}. Furthermore, \cite{FLARES-VII} showed that all \flares\ galaxies are highly $\alpha$-enhanced, with the shift from solar $[\mathrm{O/Fe}] \sim 0.7 \ \rm{dex}$ on average by $z=7$. FRA-1 and FRA-2 both fall marginally below this median relation given their stellar mass, with $[\mathrm{O/Fe}] = 0.67$ and $0.69 \ \rm{dex}$ respectively. It should be noted that this does not manifest in the spectra and photometry used thus far, as the \texttt{BPASS-v2.2.1} models do not include $\alpha$-enhanced spectra.

While RQG is likely to be $\alpha$-enhanced, \flares\ does not predict it to be preferentially so relative to the general population. If caused by $\alpha$-enhancement, the metallicity degeneracy observed by \citetalias{Weibel_2025} could therefore affect the majority of measurements at high-redshift by an even greater degree. We investigate its origin by applying the \prospector\ routines adopted by \citetalias{Weibel_2025} to our simulated analogues and evaluating the inferred metallicities. This replicates the \citetalias{Weibel_2025} approach exactly, using both the spectra and photometry, identical models and priors, and mimicking wavelength dependent noise levels. We use the standard solar-abundance products derived in Section~\ref{sec:flares} as well as two additional sets of spectra and photometry generated using different SPS models. The first uses FSPS with MIST isochrones \citep{Choi_2016, Dotter_2016} and MILES spectra \citep{Miles_2006}. This also adopts a standard solar-abundance pattern and matches the model used internally by \prospector. The second uses \texttt{BPASS-v2.3} models, $\alpha$-enhanced by $0.6 \ \rm{dex}$ \citep[BPASS-$\alpha$;][]{Byrne_2022}. This is currently the maximum available, but still below the level measured in the analogues. This should therefore be considered a lower bound effect of $\alpha$-enhancement on RQG-like galaxy spectra. All three sets of observables were generated at the true snapshot redshift $z=7$, ensuring that the inferred properties can be compared directly to the truth. It should be highlighted at this stage that we do not change the SPS model used by \prospector\ at any point. The FSPS model described by \citetalias{Weibel_2025} is always used. By doing this we can investigate whether fitting an $\alpha$-enhanced spectrum with a solar-abundance model is what truly gives rise to the metallicity discrepancy observed by \citetalias{Weibel_2025}, as they propose.

Figure~\ref{fig:metal} compares the  metallicity inferred from the observables generated with each SPS model to the true mass weighted stellar metallicities from \flares. As discussed in Section~\ref{sec:physical}, we do not expect a notable difference between mass and luminosity weightings in these systems, so consider both quantities directly comparable. If the RQG metallicity discrepancy is caused solely by $\alpha$-enhancement, we would expect the metallicity of the analogues to be recovered well when their observables are generated using FSPS and standard BPASS. The underestimation should then manifest when we use BPASS-$\alpha$. This is clearly not the case, as \prospector\ systematically underestimates the truth in all scenarios. Even when fitting FSPS observables with an FSPS model, we observe a $0.5$ and $0.15 \ \rm{dex}$ underestimation of metallicity in FRA-1 and FRA-2 respectively. While the $0.6 \ \rm{dex}$ shift measured from $\alpha$-enhanced FRA-1 observables is the largest, this is still consistent with the metallicity recovered under ideal conditions. Moreover, metallicity is best recovered from $\alpha$-enhanced FRA-2 observables. It is clear that $\alpha$-enhancement alone cannot be responsible for the $1 \ \rm{dex}$ shift observed in RQG. This is not particularly surprising, as we have already demonstrated that the SED can be well recovered with solar abundance ratios. The fact that metallicity is recovered equally well from BPASS and FSPS observables indicates that a misrepresentative SPS model is also incapable of producing such a shift. There is no physical motivation for the variation in these results, so we can only assume they are dominated by scatter. 

Now that $\alpha$-enhancement has been ruled out, we consider other effects that could result in the $1 \ \rm {dex}$ metallicity underestimation in RQG. It is possible that a fundamental issue with how metallicity is inferred could be responsible. The FRA-2 values are more accurate and precise than FRA-1, hinting that the nature of the galaxy plays some part. Its strong [O\textsc{ii}] and [O\textsc{iii}] emission lines are the most obvious difference in the observables and could provide a metallicity constraint that FRA-1 and RQG lack \citep{Pagel_1979, Curti_2016, Wilkins_2023a}. To investigate this, we generate a fourth set of observables (FSPS-INTRINSIC), still matching the FSPS model of \prospector, but now without nebular emission or dust attenuation. We apply \prospector\ in the same way and also display the results in Figure \ref{fig:metal}.  If a lack of emission lines causes metallicity to be underestimated, we expect little change in FRA-1 and the accuracy in FRA-2 to drop. This is not the case and the results remain consistent with those prior. 

We observe a reversed offset in V-band dust attenuation, where the inferred values are instead overestimated. This offset reaches $>1 \ \rm{dex}$ in FRA-1, whereas FRA-2 is again closer to the truth, but still systematically overestimated. While the scatter is not consistent with Figure~\ref{fig:metal}, this indicates that an underlying degeneracy could be responsible. See Figure~\ref{fig:prop_comp} for details of this offset and the recovery of all parameters listed in Table~\ref{tab:properties}. Given FRA-1 fares far worse than FRA-2, this could also suggest an issue when measuring exceptionally low attenuations from gas poor galaxies. The age is also systematically overestimated by up to $0.3 \ \rm{dex}$, hinting that the well documented age-metallicity degeneracy may be manifesting \citep{Worthey_1994, Papovich_2001}. Strong conclusions cannot be drawn from two galaxies alone, and we defer further investigation to future work on a more representative \flares\ sample. This regardless highlights the need for stronger dust constraints and potentially calls the $\mathrm{A_{V}}$ inferred from RQG into question. 

We have demonstrated that a metallicity degeneracy alone is not sufficient evidence for strong $\alpha$-enhancement. We have also shown that the metallicity of RQG analogues is systematically underestimated by the fiducial \citetalias{Weibel_2025} SED fitting approach. With this in mind, we do not consider the RQG metallicity to be in tension with its analogues. The fiducial value is highly discrepant with the stellar mass-metallicity relation and is likely a result of this systematic effect. In contrast, the super solar value is corroborated by \flares\ and is simple to explain given our understanding of the recent histories. The lack of pristine gas halts star formation, preventing the average metallicity from being regulated by young hydrogen rich stars. The metallicity is instead dominated by stars that formed $\gtrsim300 \ \rm{Myr}$ ago, from gas enriched by stellar feedback stages.

\subsection{Number Density Implications}

\begin{figure}
 	\includegraphics[width=\columnwidth]{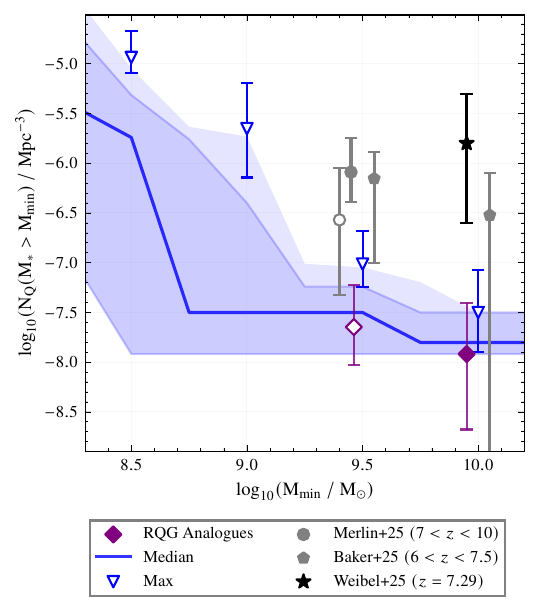}
	\caption{The number density of quiescent galaxies in \flares\ at $z=7$ exceeding a minimum stellar mass threshold $\mathrm{M_{min}}$. The blue line shows the median value recovered from combinations of systematic choices, while blue triangles show the maximum. The dark and light shaded regions indicate the $16^{\rm{th}}-84^{\rm{th}}$ and $5^{\rm{th}}-95^{\rm{th}}$ percentiles respectively. The number density implied by both RQG analogues is shown by an unfilled purple diamond and the filled diamond indicates the density implied by FRA-1 alone. Observational constraints at comparable redshift are also included. These constraints are at either $\log_{10}(\mathrm{M_{min} \ / \ \mathrm{M_{\odot}}}) = 9.5$ or 10, but shifted slightly for readability. All errorbars indicate $1\sigma$ Poisson confidence limits.
	\label{fig:density_comp}}
\end{figure}

The number density of quiescent galaxies at $z\sim7$ implied by RQG is in tension with \cite{FLARES-VIII} results from \flares\ by a factor of at least $150$. At first glance this poses a monumental challenge to galaxy formation models, but there are a number of caveats associated with identifying these sources in both regimes. The $\mathrm{sSFR}$ cut affects simulations and observations and there is no clear motivation for a specific value. The relation considered by \citetalias{Weibel_2025} adopts $\mathrm{sSFR} \ / \ \mathrm{Gyr^{-1}} < -1$, but \cite{FLARES-VIII} showed that equation \ref{eq:ssfr_cut} can produce densities $>1 \ \rm{dex}$ higher. The timescale over which quiescence is measured is not particularly important for RQG, but for the general high-redshift population this can easily move galaxies in and out of the selection region due to bursty SFHs \citep{Sun_2023, Endsley_2025, Lisiecki_2025}. This was also shown by \cite{FLARES-VIII} to produce $\sim 0.5 \ \rm{dex}$ shifts in density at the highest redshifts. The aperture size is not particularly important observationally as high-redshift galaxies are poorly resolved, but in simulations where structures are defined in three dimensional space, this can also alter the selection. Before questioning the physics of the model, we first ensure that a tension remains after considering these systematics. 

We measure the $z=7$ quiescent galaxy number density from \flares\ using every combination of six averaging timescales \mbox{$[0 - 100 \ \rm{Myr}]$}, eight apertures \mbox{$[1-100  \ \rm{kpc}]$} and three quiescence criteria \mbox{[$\mathrm{sSFR} \ / \ \mathrm{Gyr^{-1}}<\frac{0.2}{t_{\mathrm{uni}}}, -1, -2$]}. The blue line in Figure~\ref{fig:density_comp} shows the median density as a function of mass threshold across all 144 combinations and the unfilled triangles show the maximum. The minimum possible at all masses is zero and not visible. The number density of RQG-like galaxies is shown when considering both FRA-1 and FRA-2, and FRA-1 alone. The latter case implies a number density $\log_{10}(\mathrm{N_{Q}} \ / \ \mathrm{Mpc}^{-3}) = -7.92^{+0.52}_{-0.76}$. By comparing these points to the median, we see that RQG-like galaxies must be a significant population at this redshift. We therefore expect future high-redshift massive quiescent galaxy detections to be RQG-like. Tightening the mass limit intuitively has a significant effect on the number densities, but begins to diminish in importance by $\mathrm{M_{min} \ / \ \mathrm{M_{\odot}}} > 10^{9.5}$. Perhaps worryingly, the remaining mostly arbitrary choices do remain relevant in this high-mass regime at the $0.5 \ \rm{dex}$ level, and at lower masses can produce up to $\sim 3 \ \rm{dex}$ variations in density. This does not include combinations that produced zero quiescent galaxies, which are prevalent at all mass limits. This highlights the importance of not only matching the mass cut when comparing observations to simulations, but also these additional choices if constraining power is to be maximised. The fiducial \cite{FLARES-VIII} \flares\ measurement compared to by \citetalias{Weibel_2025} is consistent with our lower non-zero limit, indicating that FRA-1 was the only contributing galaxy. The gap to the \citetalias{Weibel_2025} measurement can be closed by $0.5 \ \rm{dex}$ if we adopt the maximum value presented here, but a tension of $2\sigma$ remains. This could be alleviated further by a cosmic variance term accounting for the measurement being based on a singe galaxy from a small patch of sky. These considerations have been shown to be particularly important at high redshift \citep{FLARES-X, Valentino_2023} and even more so when measuring the densities of rare objects \citep{Jespersen_2025a, Jespersen_2025b}. In contrast, the other observational constraints both come from $\sim800 \ \mathrm{arcmin}^{2}$ of sky, more than two times that of \citetalias{Weibel_2025}, and can be brought into agreement with the maximum value at the $1\sigma$ level. \cite{Merlin_2025} identify RQG as the only robust quiescent galaxy at $\mathrm{M_{min} \ / \ \mathrm{M_{\odot}}} > 10^{9.5}$, but leverage this increased area to report a $\sim0.8 \  \rm{dex}$ lower density before correcting for completeness. \cite{Baker_2025c} identify a single additional candidate at this limit, but beyond $\mathrm{M_{min} \ / \ \mathrm{M_{\odot}}} > 10^{10}$ only RQG remains. This again allows them to report a lower density. There are clearly still very few galaxies contributing to these bins and \flares\ still systematically predicts fewer sources than observed. Future observations and a full cosmic variance treatment could further alleviate these tensions, but they remain for now.

Taking these measurements at face value, the underlying physical model may need adjusting. Given that \flares\ is already capable of producing very close analogues of RQG, there is no evidence to suggest fundamental physics is missing from the model. AGN are responsible for producing these galaxies and as such may need tuning to increase the densities. All halos with mass greater than $10^{10} \ h^{-1} \mathrm{M_{\odot}}$ are seeded with a $10^{5} \ h^{-1} \mathrm{M_{\odot}}$ black hole, but only two massive galaxies are fully quenched by $z=7$. Many of these galaxies will become quiescent by $z=5$ \citep{FLARES-VIII}, but a modification to AGN feedback intensity could achieve this earlier. The energy injected by the AGN \begin{equation}
    \dot{E}_{\mathrm{AGN}} = \epsilon_{\mathrm{f}}\epsilon_{\mathrm{r}}\dot{M}_{\mathrm{\bullet}}c^{2} \,, \label{eq:accr}
\end{equation} is dependent on the feedback and radiative efficiencies $\epsilon_{\mathrm{f}}, \ \epsilon_{\mathrm{r}}$ and the accretion rate $\dot{M}_{\mathrm{\bullet}}$. Increasing either of these efficiencies would systematically boost the levels of feedback, further regulating star formation at all masses. However, this would also limit the formation of massive galaxies in general, potentially reducing the quiescent galaxy counts in high mass bins and sacrificing strong agreement with the galaxy stellar mass function \citep{Weibel_2024, Shuntov_2025}. Black holes harbour an energy reservoir that grows according to equation \ref{eq:accr} and may only feedback when this energy is sufficient to heat a gas particle by at least $\Delta T_{\mathrm{AGN}} = 10^{9} \ \rm{K}$. This minimum temperature regulates the strength and cadence of AGN feedback and therefore also impacts the number of quiescent galaxies. Lower values cause the AGN to feedback small amounts more often, steadily regulating star formation. However, this could cause gas to be less effectively ejected, limiting the number of fully quiescent galaxies. In contrast, raising the threshold will produce stronger instantaneous feedback capable of expelling more gas. While this is likely to increase the rates of quiescence, the galaxy stellar mass function may again be compromised. Increasing star formation efficiencies could counteract this in both scenarios, but an effect that becomes more prevalent once a galaxy and its black hole have accrued sufficient mass may be preferable. 

The accretion rate follows the \cite{Bondi_1944} prescription $\dot{M}_{\mathrm{Bondi}} \propto \mathrm{M^{2}_{\bullet}}$, limited by the ratio of Bondi and viscous timescales and the Eddington rate $\dot{M}_{\mathrm{Edd}} \propto \mathrm{M_{\bullet}}$. These different mass dependencies cause small black holes to be limited more regularly, but massive black holes to be regulated by a larger absolute amount. Allowing super-Eddington accretion rates would preferentially heighten the intensity of feedback in massive galaxies and increase rates of quiescence. Any change that increases the volume of expelled gas will also regulate the mass of the black hole, but ensuring a higher ratio of energy is transferred thermally rather than kinetically could alleviate this \citep{Weinberger_2016, Beckmann_2017, Dave_2019}. In this regime, super-Eddington rates are capable of producing more massive black holes \citep{Massonneau_2023, Husko_2025}, potentially alleviating emerging tensions at high redshift \citep{Harikane_2023, Maiolino_2024, Juodzbalis_2025}. Among other differences, the \textsc{colibre} simulations \citep{Schaye_2025} implement accretion rates of up to 100 times Eddington and reproduce the observations considered here far more accurately \citep{Chaikin_2025}. These subgrid models are highly complex \citep{Dalla_2012, crain_eagle_2015, schaye_eagle_2015} and the overall effects of adjusting parameters can only be fully understood by running new simulations, preferably spanning a range of values. Furthermore, stellar and AGN feedback have been shown to be strongly coupled so cannot be fully evaluated in isolation \citep{Tillman_2023}.

Our results could be built upon more generally by enhancing the capabilities of the simulation. While the zoom approach allows for the probing of larger effective volumes than a standard periodic box, there are still limitations on the modelling accuracy that can be achieved. Higher mass and spatial resolutions would facilitate more detailed modelling of the gas in and around the AGN accretion disc \citep{Gabor_2014a, Byrne_2024}, better constraining the strength and cadence of feedback. To accurately model its effect on star-forming gas, this feedback should be tracked using full radiative transfer \citep{Kannan_2021, Chen_2023}. We have shown previously that heating is extremely important in rapidly regulating star formation and rejuvenation is governed by the subsequent cooling. Both could be better modelled by detailed chemical networks \citep{Richlings_2014, Raouf_2023}, accounting for the absorption and radiative properties of different elements. These make up only a small subset of potential improvements, all of which come at a computational cost. These costs are currently untenable if we wish  to identify rare objects like early massive quiescent galaxies. 

Unlike \flares, the more limited volumes of traditional periodic boxes probe fewer extreme environments and often lack even a single quiescent system at these redshifts. They are therefore unable to report number densities, bar upper limits based on assuming no evolution at higher redshifts. \citetalias{Weibel_2025} show that even under these most conservative assumptions, TNG-100 \citep{Pillepich_2018}, \textsc{Simba} \citep{Dave_2019}, \textsc{galform} \citep{Lacey_2016} and standard \eagle\ all underestimate the observed densities. A handful of direct measurements do exist at $z=7$. As mentioned above, we believe the super-Eddington accretion rates allowed within the \textsc{colibre} model could be responsible for their higher number densities. These reach $\log_{10}(\mathrm{N_{Q}} \ / \ \mathrm{Mpc}^{-3}) = -6.62^{+0.23}_{-0.54}$ when $\mathrm{M_{min} \ / \ \mathrm{M_{\odot}}} = 10^{10}$, agreeing with \citetalias{Weibel_2025} and low redshift constraints. \mbox{\textsc{Magneticum}} is similarly able to produce more high-redshift quiescent galaxies, reporting $\log_{10}(\mathrm{N_{Q}} \ / \ \mathrm{Mpc}^{-3}) = -6.78$ under the same mass limit \citep{Remus_2025}. While still underestimating \citetalias{Weibel_2025}, this is a $\sim0.7 \ \rm{dex}$ increase on \flares. However, unlike \textsc{colibre}, this comes at the cost of agreement at lower redshifts. This is shown by \cite{Baker_2025a}, who identify a factor $100$ boost applied to AGN accretion rates when accreting cold gas as the likely cause of this systematic overestimation. AGN feedback again appears important for creating these galaxies, but the \mbox{\textsc{Magneticum}} approach may need further tuning. \cite{Seeyave_2025} ran the semi-analytic models \textsc{shark} \citep{Lagos_2024} and \textsc{scsam} \citep{Yung_2024} on dark matter only versions of the same \flares\ re-simulation regions considered in this work. With $\mathrm{M_{min} \ / \ \mathrm{M_{\odot}}} = 10^{9.5}$, they measure $\log_{10}(\mathrm{N_{Q}} \ / \ \mathrm{Mpc}^{-3}) = -7.5$ and $-7.75$ respectively, largely consistent with our results based on the \eagle\ model and still in tension with observations. The authors investigate how these densities change when turning off AGN feedback and find that while the quiescent fraction of central galaxies drops to almost zero in \textsc{shark}, \textsc{scsam} is mostly unaffected. While AGN are the drivers of early quenching in most models, including \flares, this is clearly not always the case. These number densities are also affected by the same systematics quantified above. 

%% file: Sections/5_conclusion.tex
\section{Conclusions} \label{sec:conc}

We have used synthetic photometry and spectra generated by \synthesizer\ to identify \flares\ analogues of RQG, the most distant massive quiescent galaxy identified to date. The galaxies best matching the shape of its SED are FRA-1 and FRA-2, which demonstrate outstanding agreement with other observable properties. These galaxies are massive and dominated by a rapidly quenched burst, demonstrating the ability of the \flares\ model to produce such extreme galaxies even earlier than observed so far. Furthermore, the stellar mass, metallicity and recent SFR of FRA-1 closely resemble those inferred from RQG, corroborating some SED fitting results.

We use these analogues to interpret the underlying physics and find massive inflows of gas from the surrounding IGM are required to reach comparable masses so early. We find stellar feedback to be ineffective at regulating star formation over extended periods, but both galaxies harbour massive central black holes. Quenching in these systems is a result of AGN feedback, which acts rapidly by heating star forming gas. Further expulsion of the gas beyond the bounds of the subhalo reduces the gas fraction in FRA-1 from $60$ to $<1\%$ in only $\sim 150 \ \rm{Myr}$, limiting dust attenuation to negligible levels. This prevents timely rejuvenation and the now massive AGN are capable of swiftly regulating future star formation. Their luminosities are likely to be exceptionally low outside of these episodes, as minimal gas leaves little material available for accretion. Indirect observations, such as those of the gas outflow rate by \cite{Valentino_2025}, are likely the most reliable way of confirming the presence of an AGN in RQG.

Such an evolutionary history causes preferentially high metallicities, with \flares\ predicting super solar enrichment in RQG. Levels of $\alpha$-enhancement are high, but below average for the epoch. The $1 \ \rm{dex}$ discrepancy in metallicities inferred by \citetalias{Weibel_2025} under two \prospector\ modelling assumptions cannot be reproduced in these galaxies by $\alpha$-enhancement alone. A degeneracy with dust attenuation or age may be responsible, but this cannot be confirmed without a more representative sample. We echo \citetalias{Weibel_2025} in highlighting the need for sub-mm observations, to better constrain the dust properties and potentially break this degeneracy.

Systematics in quiescent galaxy identification are not capable of alleviating tensions between the number density predicted by \flares\ at $z=7$, and the value implied by RQG. These systematics introduce a $3 \ \rm{dex}$ uncertainty at lower masses, but this decreases to $0.5 \ \rm{dex}$ if $\log_{10}(\mathrm{M_{min} \ / \ \mathrm{M_{\odot}}}) > 10$.  The density in \flares\ could be increased by enhancing AGN feedback. Super-Eddington accretion rates could preferentially increase feedback strength in massive galaxies, preserving agreement with the galaxy stellar mass function. Samples from more representative areas demonstrate closer agreement, but higher number statistics are required for better constraining power. Analogues of RQG appear to constitute a significant fraction of the massive quiescent galaxy population at high redshift. Further constraints could therefore come from a similar SED matching search in archival \emph{JWST} imaging.

%% file: Sections/appendix_1.tex
\section{Matching to FSPS Photometry}\label{app:matching}

Our main analysis uses the \texttt{BPASS-v2.2.1} SPS model to generate synthetic observables for \flares\ galaxies (\S\ref{sec:flares}), as we believe it to be the most representative of the high-redshift Universe. Unlike other models, BPASS includes a contribution from binary systems, which are expected to be commonplace in all galaxies \citep{Sana_2012}. This is in addition to treatments for young stars, which are intuitively abundant at early times. However, by using FSPS in their \prospector\ modelling, \citetalias{Weibel_2025} assume that it instead is most representative. To test the impact of this differing assumption on the identification of analogues, we generate new photometry at $z_{\rm{RQG}}$ using the same FSPS model and rerun the matching. As in Section~\ref{sec:alpha}, this model uses MIST isochrones \citep{Choi_2016, Dotter_2016} and MILES spectra \citep{Miles_2006}.

Reassuringly, FRA-1 and FRA-2 are still the standout matches, being the best in $59\%$ and $20\%$ of iterations respectively. Both dissimilarity scores have increased, with FRA-2's 150 now notably higher than FRA-1's 120. The drop in FRA-2 matching rate is caused by the previously weak analogue FRA-3, becoming the best match in $11\%$ of iterations. While still significantly higher than the strong analogues, its dissimilarity score has decreased from 280 to 230. We have already shown that FRA-1 and FRA-2 are the only massive quiescent galaxies at $z=7$, so this galaxy is clearly not an overlooked analogue. We therefore conclude that using FSPS would not significantly affect our interpretation of the underlying physics.

\begin{figure}
\includegraphics[width=\columnwidth]{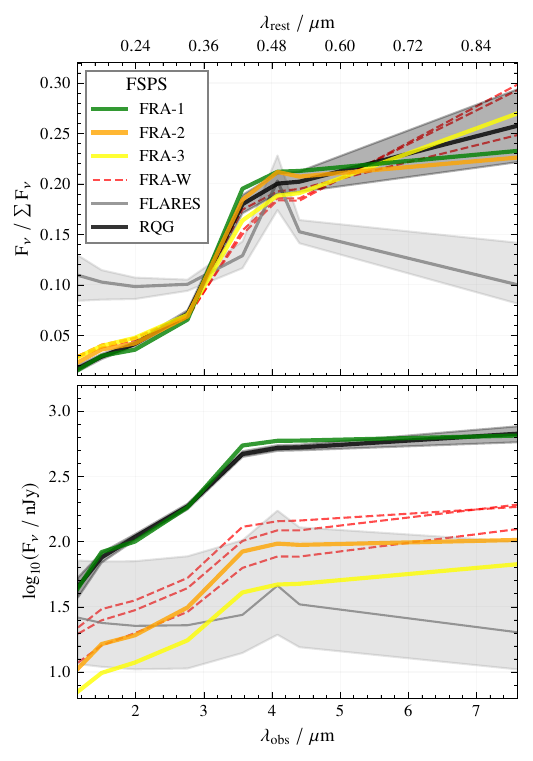}
	\caption{\emph{Top}: The total flux normalised SEDs of RQG and its potential analogues when using FSPS to generate the photometry. The colouring matches Figure~\ref{fig:seds_comp}, with the previously weak analogue FRA-3 now shown by a yellow line. Shaded regions indicate $1\sigma$ confidence limits. \emph{Bottom}: The absolute SEDs of the same galaxies. 
	\label{fig:fsps_seds}}
\end{figure}

The top panel of Figure~\ref{fig:fsps_seds} shows the normalised SEDs from FSPS, which fail to trace the shape of RQG as closely as BPASS (\S\ref{sec:matching}). No galaxy is capable of reproducing the colours at wavelengths longer than the Balmer break, with both strong analogues losing their high F444W-F770W component. If we had not already shown otherwise, these matches could appear coincidental, but we would not immediately conclude that \flares\ is incapable of producing galaxies that look like RQG. These galaxies are still red with strong Balmer breaks, so would likely pass our checks in Section~\ref{sec:observables}. The bottom panel shows that these galaxies are now intrinsically fainter, allowing FRA-1 to match RQG almost exactly. While this is desirable, we consider the shape of the SED to be more important, as this is what really encodes the properties and history of a galaxy. It has been shown that these are common between RQG and its analogues, particularly FRA-1, which should be represented by similarly shaped SEDs. We consider the ability of BPASS to achieve this as strong justification for using it as our fiducial model, in addition to its more appropriate modelling.

%% file: Sections/appendix_2.tex
\section{Prospector Derived Quantities}\label{app:prospector}

\begin{figure}
\includegraphics[width=\columnwidth]{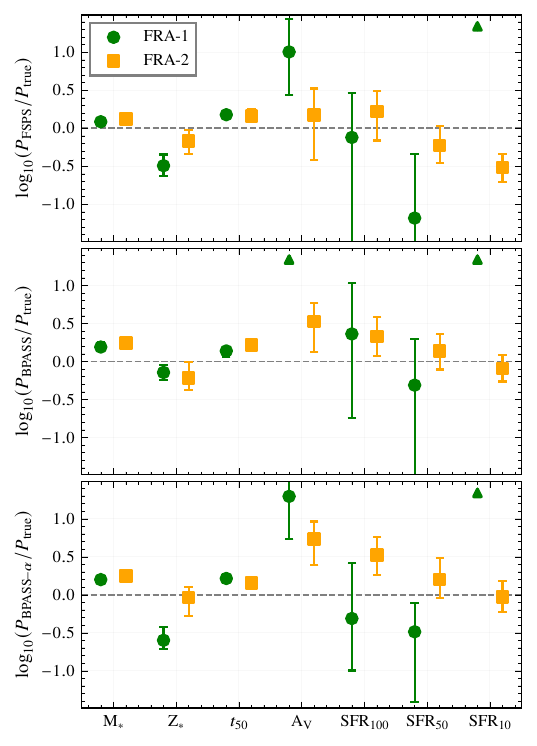}
	\caption{The ratio of \prospector\ inferred physical properties to those measured directly from \flares. The top, middle and bottom panels show the results when different SPS models are used to generate the synthetic observables. Arrowheads indicate a $>1.5 \ \rm{dex}$ shift in that direction.
	\label{fig:prop_comp}}
\end{figure}

\begin{figure}
 	\includegraphics[width=\columnwidth]{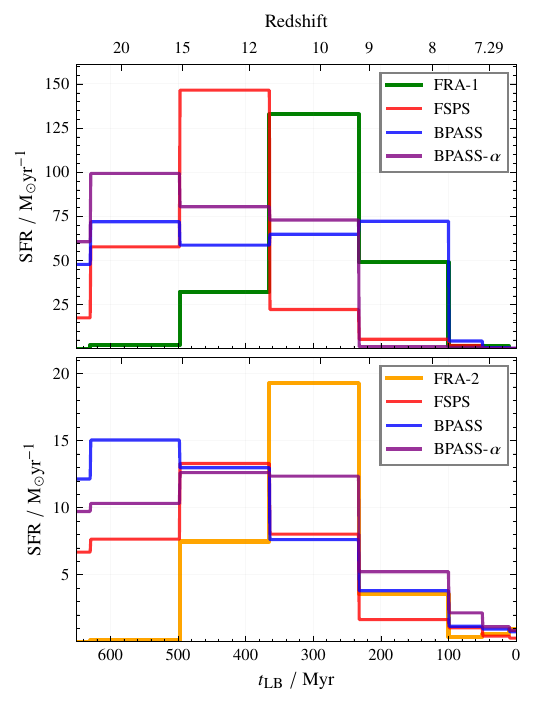}
	\caption{The absolute star formation histories of FRA-1 (top) and FRA-2 (bottom) as a function of lookback time. True histories are indicated by green and orange lines respectively. Those inferred by \prospector\ from observables generated with FSPS, BPASS and BPASS-$\alpha$ are shown by red, blue and purple lines respectively for both galaxies. Uncertainties have been omitted for readability.
	\label{fig:prospector_sfhs}}
\end{figure}

Figure~\ref{fig:prop_comp} shows the difference in physical properties derived directly from \flares\ and inferred from the photometry and spectra using \prospector. Each of the three panels show the result when a different SPS model was used to generate the observables. As highlighted in Section~\ref{sec:alpha}, the metallicity is underestimated in all scenarios, but most prevalently in FRA-1. In contrast, V-band dust attenuation is consistently overestimated by $> 1 \ \rm{dex}$ in FRA-1 and to a lower level in FRA-2. These are clearly notable deviations, although large uncertainties ensure the significance is often $<2\sigma$. As discussed in the main text, this could suggest an issue with inferring the levels of dust attenuation in gas poor galaxies, where the values are negligibly low. While not overestimated to the same degree, age appears to follow a similar trend, with the scatter between SPS models perhaps more consistent with the pattern seen in metallicity. As expected given the literature, there may be a degeneracy between all three parameters. \citetalias{Weibel_2025} employ the two-parameter dust model of \cite{Kriek_2013}, but a simpler, single parameter model may perform better under these circumstances. The mass is also overestimated by $\sim 0.2 \ \rm{dex}$ in all runs, which may be of some concern given that outshining by young stars \citep{Narayanan_2024} should not significantly affect quiescent galaxies. This is the only parameter with a clear dependence on SPS model, with the masses of both analogues best recovered from FSPS observables. Those measured from BPASS are overestimated by a greater degree, which is not surprising given that it produces intrinsically brighter galaxies. The SFRs appear to suffer from significant scatter, although shorter timescales might be better constrained. Unsurprisingly,  $\mathrm{SFR}_{10}=0$ in FRA-1 is not recovered in any run. No parameter is consistently well recovered and only mass and age appear to be shifted reasonably consistently.

We display the inferred SFHs in Figure~\ref{fig:prospector_sfhs}, plotted over the true histories measured directly from \flares. As these were measured at $z=7$, the binning differs slightly from Figure~\ref{fig:normed_sfh}. While the most recent $100 \ \rm{Myr}$ is still split into bins of width 10, 40 and $50 \ \rm{Myr}$, the remaining five bins each have width $132 \ \rm{Myr}$. No run accurately recovers the full history, with all six greatly overestimating the amount of early star formation. The resulting histories appear more constant than burst dominated in most cases, but the galaxies are always identified as quiescent. An intense burst can be inferred from FRA-1, but it appears $\sim150 \ \rm{Myr}$ earlier than the truth. FRA-2 is not inferred to have undergone a notable burst and as such the magnitude of quenching is not comparable. In contrast, the quenching timescales are similar, albeit longer in the inferred histories. There are no obvious trends with SPS model, but while we have not shown the uncertainties, multiple runs demonstrate the degree of variation in these results.